\documentclass[onecolumn,showpacs,superscriptaddress,amsmath,amssymb]{revtex4}

\usepackage{graphicx}

\usepackage{amssymb}

\usepackage{bm}
\usepackage{times}

\newcommand{\ave}[1]{\left\langle#1\right\rangle}
\newcommand{\aveqe}[1]{\left\langle#1\right\rangle_\Lambda}
\newcommand{\aveqes}[1]{\langle#1\rangle_\Lambda}
\newcommand{\scals}[2]{\langle#1,#2\rangle_{\rm s}}
\newcommand{\kb}{k_{\rm B}}
\newcommand{\QE}{\Psi^\ast}
\newcommand{\np}{n_{\rm p}}
\newcommand{\bC}{\mathbf{C}}
\newcommand{\br}{\mathbf{r}}
\newcommand{\bv}{\mathbf{v}}
\newcommand{\bF}{\mathbf{F}}
\newcommand{\bQ}{\mathbf{Q}}
\newcommand{\bT}{\mathbf{T}}
\newcommand{\bzero}{\mathbf{0}}
\newcommand{\bone}{\mathbf{1}}
\newcommand{\bkappa}{\mbox{\boldmath${\kappa}$}}
\newcommand{\bLambda}{\mbox{\boldmath${\Lambda}$}}
\newcommand{\btau}{\mbox{\boldmath${\tau}$}}
\newcommand{\bxi}{\mbox{\boldmath${\xi}$}}
\newcommand{\bsigma}{\mbox{\boldmath${\sigma}$}}

\newcommand{\bhD}{\mathbf{\hat{D}}}
\newcommand{\bhM}{\mathbf{\hat{M}}}
\newcommand{\bhOmega}{\mathbf{\hat{\Omega}}}
\newcommand{\setQ}{\{\bQ_i\}}
\newcommand{\setM}{\underline{M}}
\newcommand{\setL}{\underline{\Lambda}}
\newcommand{\naq}[1]{\nabla_{\!\bQ_{#1}}}
\newcommand{\gradQ}{\nabla_{\!\bQ}}

\begin{document}



\title{Canonical Distribution Functions in Polymer Dynamics: 
I. Dilute Solutions of Flexible Polymers}


\newcommand{\TUB}{Institut\ f\"ur\ Theoretische\ Physik, 
Technische Universit\"at Berlin,
Hardenbergstr.~36, D-10623 Berlin, Germany}

\newcommand{\ETH}{ETH
Z\"urich, Department of Materials, Institute of Polymers, CH-8092
Z\"urich, Switzerland}

\author{Patrick Ilg}
\email[]{ilg@physik.tu-berlin.de}
\affiliation{\TUB}

\author{Iliya V. Karlin}
\affiliation{\ETH}

\author{Hans Christian {\"O}ttinger}
\affiliation{\ETH}

\date{\today}

\begin{abstract}
The quasi--equilibrium or maximum entropy 
approximation is applied in order to 
derive constitutive equations from kinetic models of polymer dynamics. 
It is shown in general and illustrated for an example 
how canonical distribution functions 
are obtained from the maximum entropy principle, how macroscopic 
and constitutive equations are derived therefrom and how these 
constitutive equations can be implemented numerically.  
In addition, a measure for the accuracy of the quasi--equilibrium 
approximation is proposed that can be evaluated while integrating 
the constitutive equations. 
In the example considered, it is confirmed that the accuracy 
of the approximation is increased by including more 
macroscopic variables. In steady elongational flow, it is found 
that more macroscopic variables need to be included above the  
coil--stretch transition to achieve the same accuracy as below. 
\end{abstract}


\pacs{83.80.Rs, 83.10.Gr, 05.20.Dd, 05.10.-a}

\maketitle

\section{Introduction}
\label{intro}
In this paper, the quasi--equilibrium or maximum entropy 
approximation is applied in order to 
derive constitutive equations from kinetic models of polymer dynamics. 
It is shown in general and illustrated for an example 
how canonical distribution functions 
are obtained from the maximum entropy principle, how macroscopic 
and constitutive equations are derived therefrom and how these 
constitutive equations can be implemented numerically.  
In addition, a measure for the accuracy of the quasi--equilibrium 
approximation is proposed that can be evaluated while integrating 
the constitutive equations. 

Polymer dynamics is a fascinating subject with important applications 
to polymer processing and turbulent drag reduction. 
Very recently, direct observation of the dynamics of individual 
DNA molecules in shear and elongational flow have become possible 
\cite{PeSmCh97,SmCh98,SmBaCh99}. 
The experimental results provide useful information to improve our 
understanding of polymer dynamics in a flow. 
Since deformed polymer molecules cause stresses that are responsible 
for the macroscopic non--Newtonian rheological behavior of polymer 
solutions, polymer dynamics in the presence of a flow field is 
also of great interest from an application point of view. 

Simple mechanical models are very helpful in this respect. 
Comparison of simulation results with experiments have shown that 
kinetic models of polymer molecules describe very well the rheological 
behavior of polymer fluids \cite{LiLa00,LiLaSr00} as well as the polymer 
conformation \cite{HuShLa00} in homogeneous shear and elongational 
flow. 
However, the use of 
these kinetic models is too time--consuming to be used in 
numerical flow calculations of polymer solutions. 
Therefore, simplified constitutive equations have been proposed in the 
literature that describe stresses that the polymers exert on the flow 
for a given deformation history 
(see for example \cite{BiWi95} and references therein). 
So far, however, 
no satisfactory constitutive equations are available that give 
accurate predictions under various flow conditions. 

Kinetic models of polymer dynamics may serve as starting point 
for the derivation of constitutive equations. 
In general, constitutive equations cannot be 
derived exactly from kinetic theory due to the so--called closure problem. 
In addition, closure approximations have serious 
effects on the behavior of the model, and sometimes even lead to 
unphysical predictions \cite{Keu97}. 
Recently, 
the idea of using a set of canonical distribution functions (CDF) for 
obtaining closures to the kinetic equation have been proposed in 
\cite{LiHaJaKeLe98,LiKeLe99}. 
Motivated by the ease of computation, 
very special choices for the CDF are 
considered in \cite{LiHaJaKeLe98,LiKeLe99}.

In the present work, CDF are introduced using the maximum entropy principle. 
This approach, also known as quasi--equilibrium approximation, 
is also used successfully in other branches of 
statistical physics \cite{BALIAN92,GORBAN84}. 
The resulting, quasi--equilibrium approximation to the 
dynamics, by construction, 
shows several desirable features like conservation of the 
dissipative nature of the dynamics and 
conservation of positive--definiteness of distribution functions. 
We show how this approximation can be used to obtain closed 
constitutive equations for polymer solutions. 

We also present 
an algorithm for the numerical implementation of the constitutive equations. 
In addition, a measure for the accuracy of the approximation is 
suggested that does not require solutions to the kinetic model and 
can therefore be used while solving the constitutive equation. 
In the companion paper \cite{IKO02}, this approach is applied to the 
kinetic theory of liquid--crystalline polymers.
See also \cite{Ilg01} for more details. 

This paper is organized as follows. 
Kinetic models of dilute polymer solutions are reviewed in Sec. \ref{models}. 
The maximum entropy principle is employed in Sec. \ref{QEA} in order 
to obtain canonical distribution functions (CDF) for polymer solutions. 
Some properties of the CDF are also discussed. 
Closed form constitutive equations are derived using CDF with  
special emphasize on so--called conformation tensor models. 
A numerical integration scheme for the constitutive equations is presented 
in Sec.~\ref{dualInt}. 
In Sec.~\ref{defect}, a measure for the accuracy of the approximate 
description of polymer dynamics with CDF is proposed. 
In Sec.~\ref{susy}, the derivation of constitutive equations using 
CDF together with the check of the accuracy is illustrated in detail 
for a kinetic model of polymer dynamics. 
Finally, some conclusions are offered in Sec.~\ref{end}.

\section{Kinetic Models of Dilute Polymer Solutions}
\label{models}
Let a linear, flexible polymer be represented by $N$ beads. 
The beads themselves represent many monomers, and are large compared 
to the solvent molecules. 
The distance between the beads is considered to be large enough so 
that it allows  the  
random walk statistics for the neglected segments in between,  
but close enough to maintain these segments in equilibrium. 
The connectivity of the chain therefore is guaranteed by an 
entropic potential associated with the random walk statistics between 
the beads. Ideal random walk statistics corresponds to a 
harmonic potential, giving raise to linear forces acting on the beads. 
The forces acting on the beads are often pictured as springs, 
and the kinetic models are termed ``bead--spring models''.

\subsection{Kinetic Equations of Polymer Dynamics}
Equations of motion for bead--spring models can be found in 
most of the textbooks on polymer kinetic theory 
(see e.g. \cite{BIRD87,OE96}). 
Let $\br_i, i=1,\ldots,N$, denote the position vector of the $i$--th 
bead. 
The internal conformation of the chain is described by the 
connector vectors $\bQ_i=\br_{i+1}-\br_i, i=1,\ldots,N-1$. 
In the classical approach, the polymer chain is exposed to 
a given homogeneous flow field 
$\bv(\br;t) = \bv_0(t) + \bkappa(t)\cdot \br$, where 
$\bkappa(t)=(\nabla_{\!\br}\bv(\br;t))^{\rm T}$ denotes 
the velocity gradient. 
Usually, incompressible flows,
${\rm Tr}\, \bkappa = 0$, are considered.
The configurational distribution function $\Psi$ at time $t$ 
is assumed to depend on the internal conformation but to be 
independent of the location of the center of mass, 
$\Psi=\Psi(\setQ;t)$, where $\setQ$ denotes the set of all 
connector vectors 
$\setQ=\{\bQ_1,\ldots,\bQ_{N-1}\}$. 
The famous diffusion equation for $\Psi$ reads
\begin{equation} \label{dt_Psi}
        \frac{\partial \Psi}{\partial t} = J(\Psi)
\end{equation}
with
\begin{equation} \label{diffusion_Psi}
        J(\Psi) = 
        - \sum_{i=1}^{N-1} \naq{i} \cdot
        \left\{ \bkappa\cdot \bQ_i \Psi \right\}
        + \frac{1}{2} \sum_{i,j=1}^{N-1} 
        \naq{i} \cdot \bhD_{ij} \cdot 
        \left\{ [\naq{j}U] \Psi + \naq{j} \Psi 
        \right\},
\end{equation}
where $\gradQ$ is the derivative with respect to $\bQ$. 
The last term describes diffusion due to the collisions of the solvent 
molecules with the beads.
Collisions keep each bead in thermal equilibrium with the solvent.
Temperature, Boltzmann's constant and friction coefficient of the beads 
are denoted by $T$, $\kb$ and $\zeta$, respectively. 
The function $U(\setQ)$ is the dimensionless potential of forces on the beads 
so that $\bF_i(\setQ)=\kb T\naq{i}U$ is the force acting on the 
$i$--the bead. 
It is generally assumed that only adjacent beads interact, 
the interaction being isotropic, so that $U$ can be written as 
\begin{equation} \label{1bodypot}
        U(\setQ) = \sum_{i=1}^{N-1} V(\bQ_i^2). 
\end{equation}
Long range interactions along the chain, such as those leading 
to excluded--volume effects might 
require some extra treatment. 
The diffusion matrix $\bhD_{ij}$ can be represented as 
$\bhD_{ij}=D(A^{\rm R}_{ij}\bone + \zeta\bhOmega_{ij})$, where 
the diffusion constant $D$ is given by $D=2\kb T/\zeta$. 
The unit matrix is denoted by $\bone$, 
and the $(N-1)\times (N-1)$ Rouse matrix is 
\begin{equation} \label{rouse_matrix}
        A^{\rm R}_{ij} = 2\delta_{ij} - (\delta_{i,j+1} + \delta_{i,j-1}).
\end{equation}
The tensors $\bhOmega_{ij}(\setQ)$ describe hydrodynamic interactions of bead 
$i$ and $j$ as the movement of the beads disturbs the flow field. 
Typically, the hydrodynamic interactions are described by (regularized) 
Oseen--Burger tensors \cite{OE96}. 
The first term on the right hand side of Eq.~(\ref{diffusion_Psi}) 
accounts for the hydrodynamic drag the beads experience 
while moving through the flow. 
It is assumed that the drag force is proportional 
to the difference of bead and solvent velocity. 

The number density of polymers $n_{\rm p}=\ave{1}$ is conserved 
by the dynamics (\ref{diffusion_Psi}).  
Here and in the sequel, brackets $\ave{\bullet}$ denote 
averages with respect to $\Psi$, 
$\ave{\bullet} = \int\! d^dQ\, \bullet\, \Psi(\setQ)$ with $d=3(N-1)$.  
Eqs.~(\ref{dt_Psi})) and (\ref{diffusion_Psi}) 
describe an ensemble of non--interacting 
bead--spring chains in the presence of a flow field. 
Therefore, these models describe the dynamics of dilute 
polymer solutions. 
The Rouse model, for example, is obtained when hydrodynamic interactions 
are neglected, $\bhOmega_{ij} = \bzero$, and harmonic potentials 
$V(\bQ^2) = H \bQ^2/(2\kb T)$ are considered, $H$ denotes the spring 
constant.
However, harmonic potentials are unrealistic, since they allow 
infinite stretching of the chain in strong flows. 
Warner's finite extensible nonlinear elastic potential (FENE), 
\begin{equation} \label{FENEpot}
        V(Q^2) = - (b/2) \ln(1-Q^2/Q_{\rm max}^2),
\end{equation}
exhibits a logarithmic divergence at the maximum length 
$Q_{\rm max}$, 
and therefore prohibits further stretching \cite{Wa72}. 
The finite extensibility parameter, 
$b=HQ_{\rm max}^2/(\kb T)$, is a dimensionless 
measure of the maximum extension. 
The Hookean spring force is recovered in the limit $b\to\infty$. 

We restrict ourselves here to finite extensibility effects 
with or without hydrodynamic interactions. 
Other effects such as excluded volume, internal viscosity, 
friction coefficient varying with extension 
or anisotropic frictional drag
can also be included in bead--spring models 
(see \cite{BIRD87} and references therein).

\subsection{Entropy Production}
In the absence of flow, the distribution function $\Psi$ relaxes to its 
unique equilibrium state, $\Psi^{\rm eq} = e^{-U}/Z$, 
due to the dynamics (\ref{dt_Psi}) and (\ref{diffusion_Psi}),  
where $Z$ is the normalization constant. 
The Boltzmann entropy functional,
\begin{equation} \label{S_Boltzmann}
        S[\Psi] = -\kb \int\! d^dQ\ \Psi(\setQ) 
        \ln[\Psi(\setQ)/\Psi^{\rm eq}(\setQ)], 
\end{equation}
is strictly non--decreasing during the relaxation 
due to the dynamics (\ref{dt_Psi}) and (\ref{diffusion_Psi}) 
for $\bkappa=\bzero$. 
The entropy production $\dot{S}$ is defined as the time evolution 
of the entropy functional (\ref{S_Boltzmann}), 
\begin{equation} \label{dotS}
        \dot{S} = \int\! d^dQ\ 
        \frac{\delta S[\Psi]}{\delta \Psi(\setQ)} J(\Psi),
\end{equation}
where $\delta/\delta \Psi$ is the Volterra functional derivative.
According to Eqs.~(\ref{diffusion_Psi}) and (\ref{dotS}), 
the entropy production can be split in two parts, 
$\dot{S} = \dot{S}^{h} + \dot{S}^{\rm d}$, 
corresponding to the first and second term on the right hand side 
of Eq.~(\ref{diffusion_Psi}), respectively. 
It is easily verified that $S$ 
is non--decreasing in the absence of flow,
\begin{equation} \label{dotS_irr}
        \dot{S}^{\rm d} = \frac{\kb}{2} \sum_{i,j=1}^{N-1} 
        \int\!d^dQ\ 
        \left( \naq{i} \frac{\delta S[\Psi]}{\delta \Psi} \right)
        \cdot\bhD_{ij}\Psi\cdot
        \left( \naq{j} \frac{\delta S[\Psi]}{\delta \Psi} \right) 
        \geq 0,
\end{equation}
if the diffusion matrix $\bhD$ is positive semi definite. 
In the presence of a flow field, the entropy exchange $\dot{S}^{h}$ 
of the polymer and the solvent subsystem is given by
\begin{equation} \label{dotS_rev}
        \dot{S}^{h} = -\frac{1}{T}\bkappa\colon\btau^{\rm p}
\end{equation}
which is not sign--definite. 
In Eq.~(\ref{dotS_rev}), the polymer contribution to the stress 
tensor $\btau^{\rm p}$ has been introduced which is given 
by Kramer's expression 
\cite{BIRD87,OE96}
\begin{equation} \label{kramers} 
        \btau^{\rm p}[\Psi] = (N-1)n_{\rm p}\kb T \bone - 
        \kb T \sum_{i=1}^{N-1} 
        \int\! d^dQ\ \Psi(\setQ) \bQ_i \naq{i} U(\setQ).
\end{equation}

\section{Canonical Distribution Functions in Polymer Dynamics}
\label{QEA}
Within kinetic theory, the state of the polymer at time $t$ is described 
by the configurational distribution function $\Psi(\setQ;t)$ and 
Eqs.~(\ref{dt_Psi}), (\ref{diffusion_Psi}) and (\ref{kramers}) specify the 
dynamic and viscoelastic behavior of dilute polymer solutions 
for a given homogeneous flow field. 
In viscoelastic flow calculations, the velocity field has to be
found numerically together with the dynamics of the polymers. 
The combined simulation of Navier--Stokes equation with the 
kinetic equation (\ref{dt_Psi}) is however 
very time--consuming. 
Therefore, there is an enormous interest in deriving 
closed constitutive equations from the kinetic equation. 
In general, this derivation cannot be done rigorously 
due to the so--called closure problem and certain approximations  
have to be invoked. 

\subsection{The Closure Problem}
It is assumed that the macroscopic behavior of the polymers can 
be described by a set of $n$ macroscopic variables, which are 
defined as functionals of the distribution function,   
$M_k=\hat{\Pi}_k[\Psi]$, $k=1,\ldots,n$. 
We consider macroscopic variables that are linear functionals (moments) 
of the distribution function $\Psi$
\begin{equation} \label{def_Mk}
        M_k(t) = \int\!d^dQ\ m_k(\setQ)\Psi(\setQ;t).
\end{equation}
The set of macroscopic variables need not be identical with the components 
of the stress tensor $\btau^{\rm p}$. 
We only require that the stress tensor (\ref{kramers}) is determined by the 
values of the macroscopic variables $\setM=\{M_1,\ldots,M_n\}$. 
From the chain rule of differentiation and Eq.~(\ref{dt_Psi}), 
the time evolution of $M_k$ is given by 
\begin{equation} \label{dt_Mk}
        \dot{M}_k = \int\!d^dQ\ m_k(\setQ)J(\Psi)
\end{equation}
which do not constitute, in general, 
a closed set of equations for the macroscopic variables.
In what will follow, we present a general approach to obtain closures 
of Eqs.~(\ref{dt_Mk}) based on the maximum entropy principle 
and corresponding canonical distribution functions.

\subsection{Maximum Entropy Principle and Canonical Distribution Functions} 
The reduced or macroscopic description of the system is reasonable 
if a separation of time scales of the relaxation times holds. 
The equilibrium distribution function $\Psi^{\rm eq}=e^{-U}/Z$ 
is obtained from the maximum of the entropy functional (\ref{S_Boltzmann}) 
subject to fixed normalization, $1=\int\!d^dQ\ \Psi(\setQ)$.  
The so--called generalized canonical distribution functions $\QE$ 
are obtained by the maximum entropy principle under fixed constraints 
\cite{GIBBS60,JAYNES83,GORBAN84,ZuMoRo97},
\begin{equation} \label{MaxEnt}
        S[\Psi] \to {\rm max},\qquad 
        M_k = \int\!d^dQ\ m_k(\setQ)\Psi(\setQ).
\end{equation}
Solution to the problem Eq.~(\ref{MaxEnt}), if it exists, is 
unique for concave entropy functionals like (\ref{S_Boltzmann}) 
and linear constraints given by Eq.~(\ref{def_Mk}).
For the present case the solution to Eq.~(\ref{MaxEnt}) 
is given explicitly as 
\begin{equation} \label{QE}
        \QE(\setQ) = 
        \Psi^{\rm eq}(\setQ)\exp[\sum_{l=1}^n \Lambda_lm_l(\setQ) 
        + \Lambda_0]
\end{equation}
and is called canonical distribution function (CDF) 
(or generalized canonical state or quasi--equilibrium distribution function 
or maximum entropy state). 
The Lagrange multipliers, 
$\setL=\{\Lambda_0,\Lambda_1,\ldots,\Lambda_n\}$, 
are determined from the constraints 
\begin{equation} \label{constr1}
        n_{\rm p} = \int\!d^dQ\ 
        \Psi^{\rm eq}(\setQ)\exp[\sum_{l=1}^n \Lambda_lm_l(\setQ) 
        + \Lambda_0],
\end{equation}
and
\begin{equation} \label{constrM}
        M_k = \int\!d^dQ\ m_k(\setQ)
        \Psi^{\rm eq}(\setQ)\exp[\sum_{l=1}^n \Lambda_lm_l(\setQ) 
        + \Lambda_0].
\end{equation}
In most cases, the dependence $\setL(\setM)$ or 
$\setM(\setL)$ is not known explicitly.  
The manifold $\Omega^\ast$ that is spanned by $\QE$ when 
the macroscopic variables take on all admissible values is 
known as quasi--equilibrium manifold or canonical distribution 
function manifold \cite{GoKaIlOe01,GoKa94}. 

The macroscopic entropy $S^\ast(\setM)$ is defined as 
the entropy $S[\Psi]$ given by Eq.~(\ref{S_Boltzmann}) evaluated 
on the CDF (\ref{QE}), $S^\ast(\setM)=S[\QE]$. 
The macroscopic entropy $S^\ast$ allows to interpret the 
Lagrange multipliers as the conjugate or dual 
to the macroscopic variables, 
\begin{equation} \label{Lconj}
        \Lambda_k = - \frac{\partial S^\ast(\setM)}{\partial M_k}. 
\end{equation}

\subsection{Macroscopic dynamics}
In what will follow, the canonical distribution function manifold
$\Omega^\ast$ is used to obtain the macroscopic 
description of the polymer dynamics. 
We assume that arbitrary initial states relax quickly towards 
the manifold $\Omega^\ast$ and further evolve in the neighborhood 
of $\Omega^\ast$. 
In this case, the CDFs that span the manifold can be considered 
as representative states in the sense that moments of the solution 
to the kinetic equation (\ref{dt_Psi}) are approximated 
accurately by the corresponding moments evaluated with the CDFs. 
The validity of this assumption is closely related to the separation 
of time scales mentioned above and the choice of the macroscopic 
variables. 
This assumption can be interpreted within the context of the method 
of invariant manifold \cite{GoKa94} as 
$\Omega^\ast$ being a meaningful approximation to the relevant 
invariant manifold. A measure of the accuracy of this approximation 
is presented in Sec.~\ref{defect}. For improvements on this 
approximation see \cite{GoKa94,GoKaIlOe01}. 
We do not provide any rigorous proofs for the results obtained in the sequel. 
However, the ability of 
measuring the accuracy of the approximation is a constructive method to 
gain confidence in the approximation scheme proposed here.

Having identified the manifold $\Omega^\ast$, the macroscopic 
time evolution is defined as 
\begin{equation} \label{dtM_QE}
        \dot{M}_k^\ast = \int\!d^dQ\ m_k(\setQ) J(\QE(\setM,\setQ)),
\end{equation}
which is the dynamics of the macroscopic variables (\ref{dt_Mk}) 
evaluated on $\Omega^\ast$. 
Inserting the right hand side of Eq.~(\ref{diffusion_Psi}) for $J$,  
Eq.~(\ref{dtM_QE}) becomes 
\begin{equation} \label{dtM_macro}
        \dot{M}_k^\ast = \bkappa^{\rm T}\colon\sum_{i=1}^{N-1} 
        \aveqe{\bQ_i\naq{i}m_k} 
        -\sum_{l=1}^n\hat{M}_{kl}\Lambda_l, 
\end{equation}
where $\aveqe{\bullet}$ denotes averages with respect to the CDFs $\QE$, 
Eq.~(\ref{QE}), 
and $\hat{M}$ is a positive semi--definite matrix, 
\begin{equation} \label{Mmatrix}
        \hat{M}_{kl} = 
        \frac{1}{2} \sum_{i,j=1}^{N-1} 
        \int\!d^dQ\ 
        (\naq{i} m_k)\cdot\bhD_{ij}\QE\cdot(\naq{j} m_l).
\end{equation}
Eq.~(\ref{dtM_macro}) together with Eqs.~(\ref{constr1}), 
(\ref{constrM}) and (\ref{Mmatrix}) 
represent the set of closed equations describing the macroscopic dynamics. 
Since the CDFs are assumed to be representative distributions, 
the constitutive relation is obtained by evaluating Eq.~(\ref{kramers})
on $\Omega^\ast$, $\btau^{\rm p \ast}(\setM) = \btau^{\rm p}[\QE]$. 
In the present case it reads 
\begin{equation} \label{kramer_macro}
        \btau^{\rm p \ast}(\setM) = -\kb T \sum_{i=1}^{N-1} \sum_{k=1}^n
        \aveqes{\bQ_i\naq{i}m_k}\Lambda_k. 
\end{equation}

From Eqs.~(\ref{Lconj}) and (\ref{dtM_macro}) it is found that the macroscopic 
entropy production $\dot{S}^\ast$ is given by 
\begin{equation} \label{dtS_macro}
        \dot{S}^\ast = 
        -\frac{1}{T}\bkappa\colon\btau^{\rm p \ast} 
        + \sum_{k,l=1}^n \Lambda_k \hat{M}_{kl} \Lambda_l.
\end{equation}
and could have been obtained by evaluating Eqs.~(\ref{dotS_irr}) and 
(\ref{dotS_rev}) on the CDF (\ref{QE}). 
In the absence of flow, the macroscopic entropy production (\ref{dtS_macro}) 
is positive semi--definite as is the underlying kinetic model, 
Eq.~(\ref{dotS_irr}). This illustrates a general property of 
the quasi--equilibrium approximation that it conserves the type of dynamics 
(see, e.g. \cite{GoKaIlOe01}).

\subsection{Conformation Tensor Models} \label{ConfTen}
In the present section, we specialize to conformation tensors as 
macroscopic variables. The corresponding conformation tensor models 
are frequently used in the literature \cite{DrEdOe99}. 
The conformation tensors $\bsigma$ are the covariance matrices of the 
distribution function
\begin{equation} \label{2nd_mom}
        \bsigma_{ij} = \ave{\bQ_i\bQ_j}, \quad i,j=1,\ldots,N-1.
\end{equation}
In our notation, conformation tensor models are defined by the 
special choice $M_k\to\bsigma_{ij}$ and $m_k\to\bQ_i\bQ_j$ in 
Eq.~(\ref{def_Mk}). Corresponding Lagrange multipliers are denoted 
by $\bLambda_{ij}$.

For conformation tensor models the time evolution equation 
(\ref{dtM_macro}) reads
\begin{equation} \label{dtsigma}
        \dot{\bsigma}_{ij} = 
        \bkappa\cdot\bsigma_{ij} + \bsigma_{ij}\cdot\bkappa^{\rm T} 
        - \sum_{k,l=1}^{n}\bhM_{ijkl}\colon\bLambda_{kl},
\end{equation}
where the matrix (\ref{Mmatrix}) now reads 
\begin{equation} \label{MmatrixConfTen}
        \bhM_{ijkl} = 
        \frac{1}{2}\aveqe{\bQ_i\bhD_{jk}\bQ_l+\bQ_i\bhD_{jl}\bQ_k+
        \bQ_j\bhD_{ik}\bQ_l+\bQ_j\bhD_{il}\bQ_k}.
\end{equation}
In Eq.~(\ref{dtsigma}) one recognizes the upper--convected 
derivative, while the last term describes relaxation mechanisms. 

For the present choice of macroscopic variables, the stress tensor 
(\ref{kramer_macro}) takes the form 
\begin{equation} \label{stress_sigLa}
        \btau^{\rm p} = -2\kb T \sum_{i,j=1}^{N-1} 
        \bsigma_{ij}\cdot\bLambda_{ji}. 
\end{equation}

\section{Numerical Integration Scheme} \label{dualInt}
In the maximum entropy principle (\ref{MaxEnt}), Lagrange multipliers 
$\setL$ occur naturally to satisfy the constraints. 
In principle, the Lagrange multipliers can be expressed as functions 
of the macroscopic variables using 
the constraints (\ref{constr1}) and (\ref{constrM}). 
In general, however, no explicit expressions $\setL(\setM)$ can be 
found. 
For these cases, we here propose a method that does not require 
knowledge of the functions $\setL(\setM)$

\subsection{Dual Variables and Their Dynamics} \label{DualDyn}
A change of the independent variables from the set $\setM$ to $\setL$ 
is done by Legendre transform. The Legendre transform of the 
macroscopic entropy $S^\ast(\setM)$ is the generating function 
$G(\setL)=S^\ast(\setM)+\sum_k\Lambda_kM_k$, which can be written as 
\begin{equation} \label{generatG}
        G(\setL) = \ln \int\!d^dQ\ 
        \Psi^{\rm eq}(\setQ) \exp[\sum_{l=1}^n \Lambda_lm_l(\setQ)]. 
\end{equation}
The macroscopic variables and higher order moments can be obtained 
by derivatives of $G$ with respect to $\Lambda$, 
\begin{equation} \label{Mk_L}
        M_k(\setL) \equiv \aveqe{m_k} = 
        \frac{\partial G(\setL)}{\partial \Lambda_k}
\end{equation}
and
\begin{equation} \label{Ckl_L}
        C_{kl}(\setL) \equiv \aveqe{m_km_l} - \aveqe{m_k}\aveqe{m_l} = 
        \frac{\partial^2 G(\setL)}{\partial \Lambda_k\partial \Lambda_l}.
\end{equation}

The Lagrange multipliers $\setL$ are now considered as primary independent 
variables that parameterize the CDFs (\ref{QE}) and that determine the 
values of the macroscopic variables (\ref{Mk_L}). 
From Eqs.~(\ref{Mk_L}) and (\ref{Ckl_L}) it is found that  
the time evolution of $\setL$ is given by
\begin{equation} \label{dtLambda}
        \dot{\Lambda}_k = 
        \sum_{l=1}^n C^{-1}_{kl}(\setL) \dot{M}_l(\setL),
\end{equation}
where $\dot{M}_k(\setL)$ is given by Eq.~(\ref{dtM_QE}). 
Eq.~(\ref{dtLambda}) is the closed macroscopic equation written in 
dual variables. 
Note, that the constitutive relation (\ref{kramer_macro}) 
is already given as a function of $\setL$.

\subsection{Numerical Integration Scheme}
\label{DualInt}
The reformulation of the macroscopic dynamics (\ref{dtM_macro}) 
in terms of the dual variables $\setL$ described in Sec.~\ref{DualDyn} 
is suitable for numerical implementation. 
The Lagrange multipliers $\setL$ now play the role of independent 
dynamic variables, instead of $\setM$. 
To advance given values $\setL(t)$ at time $t$ to their values 
$\setL(t+\tau)$, with small time step $\tau$, the following first order 
integration scheme is proposed \cite{IKO00b}: 
\begin{enumerate}
\item The new values of the macroscopic variables $\setM(t+\tau)$ 
are found from the values $\setL(t)$ by 
\begin{equation} \label{dualInt1}
  \frac{M_k(t+\tau)-M_k(t)}{\tau} = 
  \int\!d^dQ\ m_k(\setQ) J(\QE(\setL(t),\setQ)),
\end{equation}
where $J(\QE(\setL(t),\setQ))$ denotes the right hand side of 
Eq.~(\ref{diffusion_Psi}), evaluated with the CDF (\ref{QE}) 
with $\setL(t)$ the actual values of the Lagrange multipliers. 
\item The matrix $C_{kl}$, defined in Eq.~(\ref{Ckl_L}), is evaluated 
from 
\begin{equation} \label{dualInt2}
  C_{kl}(t) = \aveqe{m_km_l}(t) -M_k(t)M_l(t),
\end{equation}
where $\aveqe{m_km_l}$ is calculated from Eq.~(\ref{dualInt1}), 
where $m_k$ is now replaced by $m_km_l$. 
\item The $n\times n$ matrix $\bC(t)$ is inverted numerically to 
give $C^{-1}(t)$.   
\item The values of the Lagrange multipliers, $\setL(t+\tau)$, 
are given by 
\begin{equation} \label{dualInt3}
  \frac{\Lambda_k(t+\tau)-\Lambda_k(t)}{\tau} = 
  \sum_{l=1}^nC^{-1}_{kl}(t) \frac{M_k(t+\tau)-M_k(t)}{\tau}
\end{equation}
\end{enumerate}
This concludes one time step of integration. 
The integration scheme has to be supplemented by initial conditions 
$\setL(0)$. The special case of equilibrium initial conditions 
corresponds to $\setL(0)=\underline{0}$. 

Note, that evaluating the matrix (\ref{Ckl_L}) requires moments of 
$\Psi$ which are of higher order than the macroscopic variables 
themselves. If $n$ denotes the number of macroscopic variables, 
the numerical integration scheme requires in total $n(n+3)/2$ 
integrals per time step to evaluate $\setM$ and $\bC$. 
Due to the symmetry of $\bC$, this number is of order $n^2/2$ rather 
than $n^2$. The evaluation of $Y_k$ may require additional integrals. 
The evaluation of all these integrals with standard numerical 
methods might be time--consuming, 
especially for high--dimensional integrals. 
It is demonstrated in \cite{HuBa98} that under certain circumstances 
these integrals can be evaluated
efficiently by adapting methods of fast Fourier transformations.

\section{Accuracy of Approximation by Canonical Distribution Functions}
\label{defect}
The reduced description outlined in Sec.~\ref{QEA}
obviously imposes restrictions or approximations to the underlying 
kinetic model. 
Therefore, the accuracy of the reduced description needs to be discussed. 
The convergence of arbitrary initial conditions to the manifold 
$\Omega^\ast$ is of obvious importance, but is in general difficult to 
study. 
For a concrete example, the convergence or non--convergence is 
investigated in \cite{IlKa00a}.

CDFs are in general not solutions to the kinetic equation 
(\ref{dt_Psi}), i.e.~the canonical distribution function 
subspace $\Omega^\ast$ is not dynamically invariant.
The dynamic variance $\Delta$ is defined as 
\begin{equation} \label{def_defect}
        \Delta(\setM) = 
        J(\QE) - \sum_{k=1}^n\frac{\partial \QE}{\partial M_k}\dot{M}_k, 
\end{equation}
which is the difference between the microscopic time evolution evaluated 
on $\Omega^\ast$ and the time evolution of the CDFs due to the macroscopic 
dynamics. 
The dynamic variance $\Delta$ can be considered as `defect' of the 
macroscopic description. 
While the method of invariant manifold \cite{GoKa94} uses $\Delta$ 
to obtain improved manifolds, we will later use $|\Delta |$ as a measure of 
the accuracy of the approximation by CDFs. 
The norm $|\bullet |$ used later is related to $\Delta_\ell$, which is 
the difference of the dynamics of moment $M_\ell$ due to the microscopic 
and macroscopic dynamics,  
\begin{equation} \label{Delta_ell}
        \Delta_\ell \equiv \int\!d^dQ\ m_\ell \Delta = 
        \frac{d}{dt}\aveqes{m_\ell} - 
        [\aveqes{m_\ell m_j}-\aveqes{m_\ell}\aveqes{m_j}] 
        (C^{-1})_{j k}\dot{M}_k.
\end{equation}
$\Delta_\ell$ vanishes identically for $\ell \leq n$ as it has to. 
For $\ell > n$, the quantities $\aveqes{m_\ell}$ are not included 
in the list of macroscopic variables (\ref{MaxEnt}). If these 
quantities are defined such that the set 
$\{\setM,\aveqes{m_\ell}\}$ is linear independent of $\setM$, 
Eq.~(\ref{Delta_ell}) provides valuable information on the dynamics 
of variables that are not included in the list of macroscopic variables. 

Inserting in Eq.~(\ref{def_defect}) the right hand side of 
Eq.~(\ref{diffusion_Psi}) for $J$ 
and Eq.~(\ref{QE}) for $\QE$ together with the time evolution 
of the macroscopic variables (\ref{dtM_macro}) one obtains 
$\Delta_{\setM} = \QE\hat{\Delta}_{\setM}$ with 
\begin{eqnarray} \label{varBS}
        \lefteqn{\hat{\Delta}_{\setM} =  - {\rm Tr}\, \bkappa - 
        \sum_{i=1}^{N-1} \bQ_i\cdot\bkappa^{\rm T}\cdot(-\naq{i}U+\bxi^0_i) 
        - \sum_{i,k=1}^{N-1} \bkappa\colon\aveqe{[\naq{i}m_k]\bQ_i} \xi^1_k }
        \\
        & &{} 
        + \frac{1}{2} \sum_{i,j=1}^{N-1} \naq{i}\cdot\bhD_{ij}\cdot\bxi^0_j 
        -\frac{1}{2} \sum_{i,j,k=1}^{N-1} \left(
        \aveqe{\naq{j}\bhD_{ij}[\naq{i}m_k]} 
        -\aveqe{[\naq{i}m_k]\bhD_{ij}\naq{j}U} \right) \xi^1_k. 
        \nonumber
\end{eqnarray}
In Eq.~(\ref{varBS}), we have defined 
$\bxi^0_i=\sum_k\Lambda_k\naq{i}m_k$ and 
$\xi^1_k=\partial \ln \QE/\partial M_k$.
For the special case of conformation tensors as the macroscopic variables
the expression (\ref{varBS}) simplifies on the replacement 
$\setM\to\bQ\bQ$. 
If in addition one specializes to certain mean--field potentials, then 
the expression for the variance becomes most transparent. 
This is demonstrated in the sequel.

\subsection{Gaussian Manifolds of Conformation Tensor Models}
In this section we consider general potentials of the form 
\begin{equation} \label{U^P}
        U(\{ \bsigma_{ij} \}) = 
        \sum_{i=1}^{N-1} V({\rm Tr}\, \bsigma_{ii})
\end{equation} 
for the kinetic equation (\ref{dt_Psi}). 
The restriction to these potentials is very important. They can either 
be viewed as mean--field approximations to an underlying potential 
of the form (\ref{1bodypot}) or as models in their own right. 
Concerning the first interpretation, a possible derivation together 
with systematic corrections to this approximation is illustrated 
in \cite{GoKaIlOe01}. 

For conformation tensor models, the equilibrium states are isotropic 
Gaussian distributions if the potentials are of the form (\ref{U^P}). 
An important property of conformation tensor models for these potentials 
is that also the quasi-equilibrium states (\ref{QE}) 
are Gaussian.
In this particular case, Eq.~(\ref{2nd_mom}) can be evaluated 
by Gaussian integration and subsequently be inverted to 
give the Lagrange multipliers as explicit functions of the 
macroscopic variables
\begin{equation} \label{lambda_sigma}
        \bLambda_{ij}(\{\bsigma_{ij}\}) = V'({\rm Tr}\, \bsigma_{ii})
        \bone\delta_{ij} - 
        \frac{1}{2}(\bsigma^{-1})_{ij}.
\end{equation}
Inserting the functional form of the Lagrange multipliers 
(\ref{lambda_sigma}) and the potential (\ref{U^P}) 
into the quasi--equilibrium distribution function we obtain 
the standard representation of the symmetric Gaussian distribution functions 
\begin{equation} \label{3NGauss}
        \Psi_{\bsigma}(\setQ) = \frac{1}{\sqrt{(2\pi)^d \det{\bsigma}}} 
        \exp \left\{-\frac{1}{2} \sum_{i,j=1}^{N-1} \bQ_i\cdot 
        (\bsigma^{-1})_{ij}\cdot\bQ_j \right\}. 
\end{equation}
Up to an irrelevant constant, 
also the quasi-equilibrium entropy 
$S^\ast(\bsigma)$ is found from Eq.~(\ref{lambda_sigma}) 
as explicit function of $\bsigma$,
\begin{equation} \label{S_QEA_sigma}
        S^\ast(\{\bsigma_{ij}\}) = \frac{1}{2}\ln {\rm det}\ \bsigma - 
        U(\{\bsigma_{ij}\}). 
\end{equation}
The first term on the right hand side of Eq.~(\ref{S_QEA_sigma}) 
originates from the normalization of $\Psi_{\bsigma}$ 
while the second term 
contains the entropic spring potential.

\subsection{Variance of Conformation Tensor Models} 
\label{invarianceBS}
For potentials of the form (\ref{U^P}), 
the time evolution equation for the conformation tensor 
$\bsigma$ is obtained 
from Eqs.~(\ref{dtsigma}), (\ref{MmatrixConfTen}) 
and (\ref{lambda_sigma}) and reads
\begin{equation} \label{dotQQ_BeadSpring}
        \dot{\bsigma}_{ij} = 
        \bkappa\cdot\bsigma_{ij} + \bsigma_{ij}\cdot\bkappa^{\rm T} -
        \sum_{k=1}^{N-1} V'({\rm Tr}\, \bsigma_{kk}) 
        \aveqes{\bQ_i\bhD_{jk}\cdot\bQ_k+\bhD_{ik}\cdot\bQ_k\bQ_j}
        + \aveqes{\bhD_{ij}}, 
\end{equation}
where use has been made of the decomposition of moments for the Gaussian 
distribution functions (see e.g. Appendix A of \cite{Oe88}). 
It is not surprising, that 
Eq.~(\ref{dotQQ_BeadSpring}) coincides with the result of the 
Gaussian approximation used in \cite{Oe88}, the averages being performed with 
the Gaussian distribution (\ref{3NGauss}). 
Eq.~(\ref{dotQQ_BeadSpring}) is an obvious generalization of 
Eq.~(5) in \cite{Oe88} for the anharmonic potentials of the form (\ref{U^P}). 
In Eq.~(\ref{dotQQ_BeadSpring}) the ``incompressibility'' condition 
$\nabla\cdot\bhD = \bzero$ for hydrodynamic interaction 
tensors is assumed. 

The variance of the QEA (\ref{varBS}) can be calculated 
for the kinetic equation (\ref{dt_Psi}) by Gaussian 
integrals. 
The result is
\begin{equation} \label{varianceBeadSpring}
        \Delta_{\bsigma} = \Psi_{\bsigma} [
        -\sum_{i,j=1}^{N-1}{\rm Tr} 
        \left(\bT_{ij}\cdot(\bsigma^{-1})_{ji}\right) 
        + \sum_{i,j,k,l=1}^{N-1}((\bsigma^{-1})_{ik}\cdot\bQ_k)^{\rm T}\cdot
        \bT_{ij}\cdot((\bsigma^{-1})_{jl}\cdot\bQ_l)  ],
\end{equation}
where we have introduced the tensor $\bT_{ij}=\bT^0_{ij}+\bT^1_{ij}$, 
with the contribution of the Brownian part as 
$\bT^0_{ij} = \bhD_{ij} - \aveqes{\bhD_{ij}}$ and the spring contribution
as $\bT^1_{ij}$,
\begin{equation}
        \bT^1_{ij} = 
        \sum_{k=1}^{N-1} V'({\rm Tr}\, \bsigma_{kk}) 
        [\aveqes{\bhD_{il}\cdot\bQ_l\bQ_j} - 
        \bhD_{il}\cdot\bsigma_{lj}]. 
\end{equation}
We will see in a moment why the flow term does not contribute to 
$\Delta_{\bsigma}$.

In the absence of hydrodynamic interactions or for 
pre-- or self--consistently averaged hydrodynamic interactions, 
the diffusion matrix $\bhD$ is independent of $\setQ$.  
Upon inspection of the variance (\ref{varianceBeadSpring}), 
one notices it immediately that in these cases  
the variance $\Delta_{\bsigma}$ vanishes identically. 
In these case,
the drift term in the diffusion equation (\ref{diffusion_Psi})
is linear in $\bQ$ and the diffusion term is independent of $\bQ$. 
In this case, Gaussian distributions provide 
exact solutions to the diffusion equation 
\cite{Oe87}. 
Examples are the Rouse and Zimm models but also models that incorporate 
finite extensibility effects in the Peterlin approximation. 
From the above discussion we learn that in this case the 
Gaussian distributions constitute the exact invariant manifold of the 
dynamics (\ref{dt_Psi}) and Eq.~(\ref{dotQQ_BeadSpring}) 
represents the time evolution equation on the invariant manifold. 
Since there exist many solutions to Fokker-Planck equation the 
relevance of the Gaussian solutions is discussed in the next 
section. 

For general hydrodynamic interactions the variance 
(\ref{varianceBeadSpring}) does not vanish and the quasi-equilibrium 
approximation (\ref{3NGauss}) is no longer exact. 
It may serve as initial approximation as, for example in 
\cite{Oe88}. 
Possible improvements on the QEA can be found in \cite{GoKaIlOe01}.

\section{Illustration}
\label{susy}
As we have already mentioned, the choice of macroscopic variables is not 
unique. Very different choices of macroscopic variables are 
possible, and correspond to 
different sets of microscopic states accessible to the 
macroscopic description. 
In the literature, 
conformation tensors are frequently used as macroscopic variables. 
In Sec.~\ref{invarianceBS}, it is shown for a particular class of 
models, that the quasi--equilibrium 
manifold for these variables is the invariant manifold. 
In case the invariant manifold is stable, one observes a fast 
relaxation to the invariant manifold followed by a slow evolution on the 
manifold. 

In general, the quasi--equilibrium manifold is not invariant and 
improvements on this approximation become an important issue. 
If the set of macroscopic variables is maintained, 
successive improvements on the QEA can be found along the lines 
of the method of invariant manifolds \cite{GoKa94}. 
Another option is to extend the set of macroscopic variables and 
apply the QEA for the larger set of variables. 
Here, we illustrate the latter option for the kinetic model 
of polymer dynamics considered in \cite{IlKaSu00}. 

In the context of polymer dynamics, the problem of the choice of 
macroscopic variables is encountered frequently. 
Recent experiments of Chu and co--workers \cite{PeSmCh97} clearly 
demonstrate the dependence of the polymer dynamics in flow on the 
initial state. 
A possible classification of initial states into dumbbell, 
half--dumbbell, folded, coiled and kinked conformations is proposed in 
\cite{LiLaSr00}.
The failure of the macroscopic description in terms of a single 
conformation tensor has attracted considerable interest and 
led de Gennes to the notion of `molecular individualism' 
\cite{Ge97}.

In order to deal with this situation, the macroscopic description 
requires an extended set of variables. 
In \cite{LiHaJaKeLe98,LiKeLe99}, the addition of one extra variable 
associated to the fourth moment of the distribution function 
is proposed. 
In \cite{LiHaJaKeLe98,LiKeLe99}, also a closure approximation for this 
set of variables is proposed. 
In this closure, however, a penalty function has to 
be introduced ad hoc in order to obtain physically admissible 
results. 

Here, we study the macroscopic description for an extended set of 
variables in the QEA. 
As demonstrated in \cite{GORBAN84}, 
the QEA is well--behaved for any choice of macroscopic variables. 
In the limit where the number of macroscopic variables goes to 
infinity, the microscopic description is recovered since no 
variation over microscopic states is performed in the maximum entropy 
principle and therefore no reduction 
of description is associated with the QEA. 
We illustrate here for a finitely extensible dumbbell model 
that the macroscopic description converges fast with increasing 
number of macroscopic variables. 
We consider the one--dimensional model proposed in \cite{IlKaSu00}, 
with the spring force
\begin{equation} \label{1Dforce}
        F(Q)=HQ_1\tan(Q/Q_1),\quad \mbox{for}\ 
        -Q_{\rm max} < Q < Q_{\rm max}
\end{equation}
where $Q_1=2Q_{\rm max}/\pi$ is determined by the maximum extension 
$Q_{\rm max}$ and 
$H$ denotes the ``spring constant'' in the sense $F(Q) \to HQ$ for 
$Q\to 0$.  
The force (\ref{1Dforce}) approximates the inverse Langevin function 
very well and allows the exact calculation of the 
spectrum of relaxation times of the associate Fokker--Planck 
operator. The result is 
\begin{equation} \label{1Dvalues}
        \lambda_n = 2\lambda_{\rm H} \left[n+n^2/b_1 \right]^{-1}, \quad 
        n=1,2,\ldots 
\end{equation}
where $\lambda_{\rm H}=\zeta/(4H)$ is the time constant of the Hookean 
dumbbell and 
$b_1=HQ_1^2/\kb T$ a dimensionless finitely extensibility 
parameter \cite{IlKaSu00}. 
Also the exact eigenfunction have been found in \cite{IlKaSu00} 
to be given by $\Psi_n=\Phi_n\Psi_{\rm eq}^{1/2}$ with 
\begin{equation} \label{1Deigen}
        \Phi_n(Q) = C_na^\dag_0a^\dag_1\ldots a^\dag_{n-1} 
        \cos^{n+b_1/2}(Q/Q_1)
\end{equation}
where $a^\dag_k=-\nabla_{\!Q}+f_{\alpha_k}$ are generalized creation 
operators. $C_n$ denote normalization constants and 
$f_{\alpha_k}=\alpha\tan(Q/Q_1)$.
The first eigenfunctions read: $\Psi_0=\Psi_{\rm eq}$, 
$\Psi_1=N_1\sin(Q/Q_1)\Psi_{\rm eq}$, 
$\Psi_2=N_2[ (b_1+1)\sin^2(Q/Q_1)-\cos^2(Q/Q_1) ] \Psi_{\rm eq}$ 
with 
$N_1^2=b_1+2$ and $N_2^2=(b_1+4)/[2(b_1+1)]$.  
Note, that the eigenfunctions are orthonormal, 
$\scals{\Psi_n}{\Psi_m} = \delta_{nm}$, with respect to the scalar product
\begin{equation} \label{scalar}
        \scals{g}{h} = \int_{-Q_{\rm max}}^{Q_{\rm max}} 
        \!dQ\ \Psi_{\rm eq}^{-1}(Q)g(Q)h(Q). 
\end{equation}

As macroscopic variables, choose the first $n$ even dimensionless 
moments of the distribution function 
$M_k = \aveqes{\tilde{Q}^{2k}}$, $k=1,\ldots,n$ and $\tilde{Q}=Q/Q_1$ is 
the reduced extension. 
The benefit of considering a one--dimensional model is of course 
that there is no tensorial structure associated with these moments. 
The QEA for this choice of macroscopic variables is of the form 
(\ref{QE}) and reads 
\begin{equation} \label{psi_QEAsusy}
        \Psi^\ast_\Lambda(Q) = (1/z_{\rm eq})\cos^{b_1}(\tilde{Q}) 
        \exp{[\sum_{j=1}^n\Lambda_j\tilde{Q}^{2j}+\Lambda_0]}.
\end{equation}
The dynamic equations for the macroscopic variables found 
from Eq.~(\ref{dtM_macro}) are 
\begin{equation} \label{dotA_susy}
        \dot{M}_k = 2k\kappa M_k + k(2k-1)M_{k-1}/(b_1\lambda_{\rm H}) 
        - kR_k/\lambda_{\rm H}
\end{equation}
where 
$R_k=\aveqes{\tan(\tilde{Q})\tilde{Q}^{2k-1}}$. 
The polymer contribution to the stress found from Kramer's expression 
(\ref{kramers}) is 
$\tau^{\rm p}/(n_{\rm p}\kb T) = 1 - b_1R_1$. 

For given values of the $n$ Lagrange multipliers at time $t$, 
$\setL(t)$, the corresponding values of the $n$ macroscopic variables
are found from 
\begin{equation} \label{Mk_susy}
        M_k(t) = \frac{2e^{\Lambda_0(t)}}{z_{\rm eq}}
        \int_0^{\pi/2}\!d\tilde{Q}\, \tilde{Q}^{2k}
        \cos^{b_1}(\tilde{Q})\exp[\sum_{j=1}^n\Lambda(t)\tilde{Q}^{2j}]
\end{equation}
Thanks to the finite extensibility of the model, the integration 
in Eq.~(\ref{Mk_susy}) is performed over a closed interval only. 
In addition, due to the symmetry of the integrand, the integration 
is performed over the positive range of $\tilde{Q}$ only. 
The matrix $C_{kl}=\aveqes{\tilde{Q}^{2(k+l)}}$ 
is obtained by the $n(n+1)/2$ corresponding integrals. 
In this particular case, some of these integrals are identical to 
components of $\setM$. 
Inversion of the matrix $\bC$ is facilitated by the fact that it is 
a symmetric, positive semi--definite matrix. 
In total, one time step of integration requires $n(n+5)/2$ 
one--dimensional integrals similar to Eq.~(\ref{Mk_susy}).
The numerical integrations were performed using standard integration 
schemes such as Gauss--Legendre or Romberg integration 
\cite{Recipes}. 
The algorithm, described above and in Sec.~\ref{DualInt} 
was implemented and run for fixed time step $\tau$ in the order of 
$10^{-2}\ldots10^{-4}\lambda_{\rm H}$. 

In particular, we consider 
equilibrium initial conditions ($\Lambda_k(t=0)=0$) 
and steady flows and compare the quasi--equilibrium 
dynamics for different numbers $n$ of macroscopic variables. 
We generally expect the QEA to be a good approximation near 
the true equilibrium. As noted in Sec.~\ref{ConfTen}, the 
approximation (\ref{psi_QEAsusy}) is exact in the steady state 
since the one--dimensional analogue of the conformation tensor 
is kept as macroscopic variables and any flow in one dimension 
is a potential flow. 
Therefore, we expect (\ref{psi_QEAsusy}) to be a good approximation 
also near the steady state. 

\begin{figure}[h]
        \setlength{\unitlength}{1cm}
        \begin{picture}(6,6)
        \put(-10,1){\centerline{\includegraphics[width=6cm]{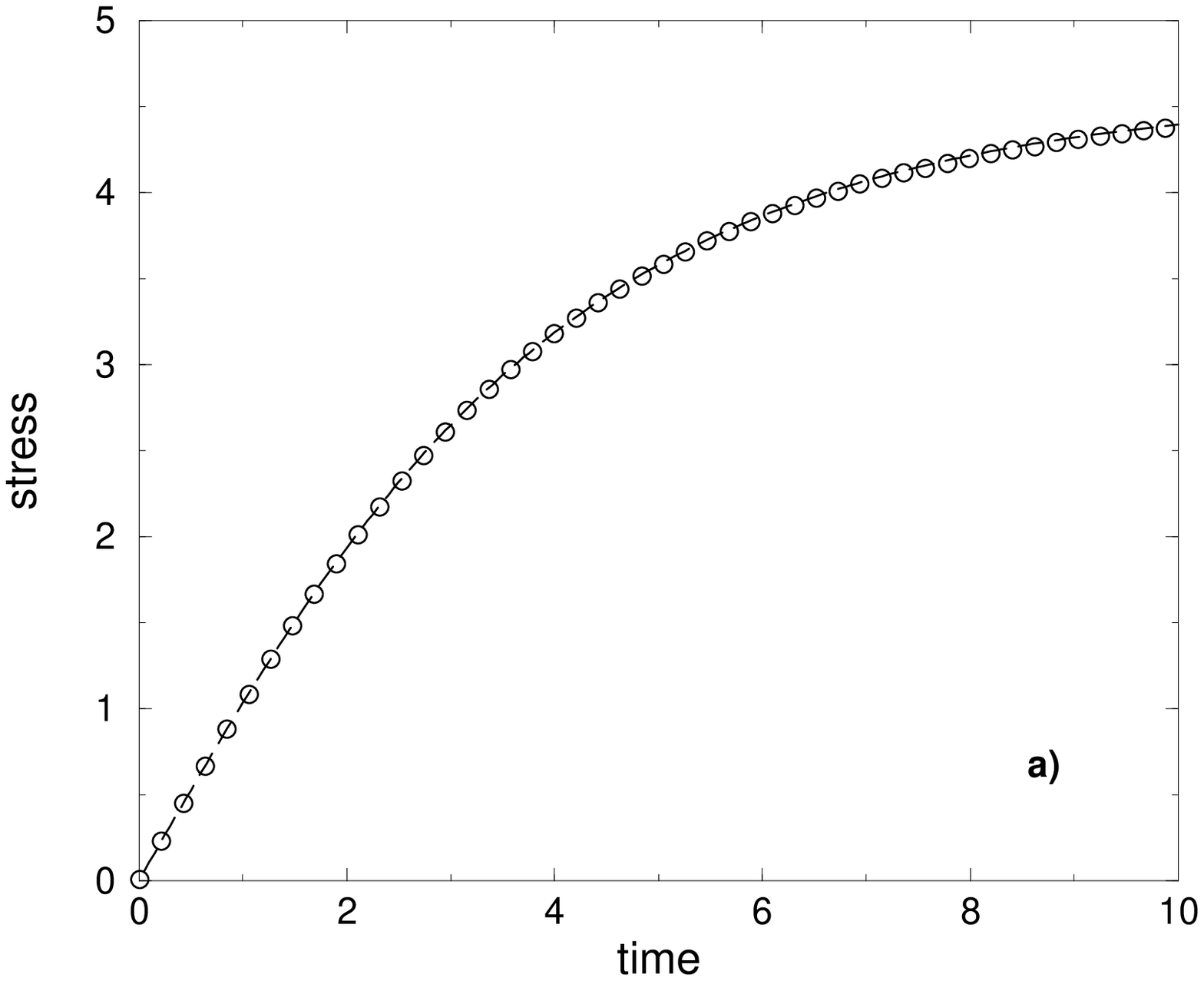}}}
        \put(-3,1){\centerline{\includegraphics[width=6cm]{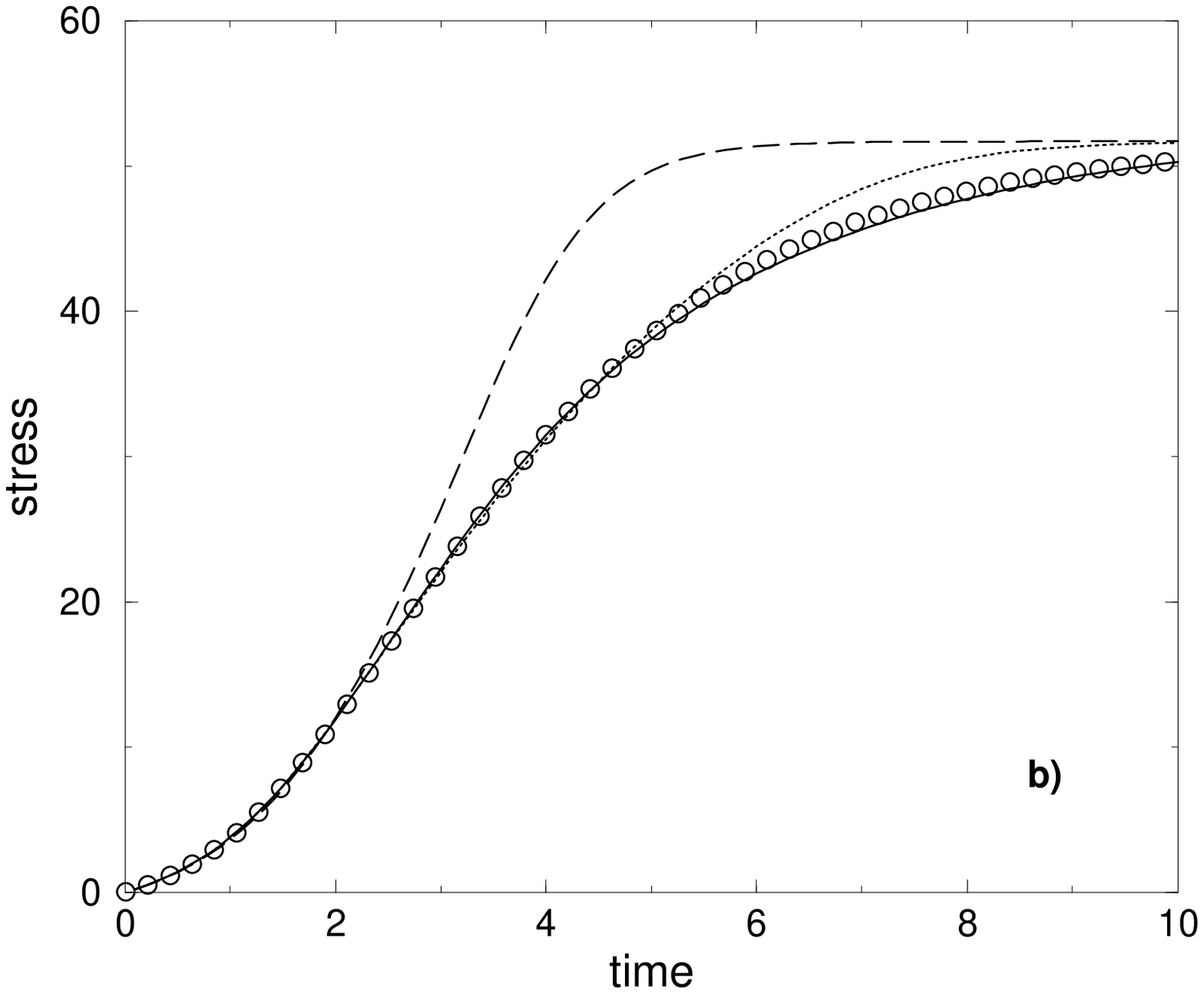}}}
\end{picture}
     \caption[QEA for SUSY] { \label{QEAsusy_fig} 
        Performance of QEA in steady elongational flow with 
        $\hat{\kappa}=0.5$ 
        $\mbox{\rm a)}$ and $\hat{\kappa}=1.0$ $\mbox{\rm b)}$. 
        Shown is the polymer stress $\tau^{\rm p}$ in units of 
        $\np\kb T$ as a function of dimensionless time 
        $t/\lambda_{\rm H}$ for 
        different sets of macroscopic variables. 
        Dashed: $n=1$, dotted: $n=2$, solid: $n=3$ and 
        symbols: Brownian dynamics simulations of full kinetic model.}
\end{figure}

In Fig.~\ref{QEAsusy_fig}, the polymer stress 
$\tau^{\rm p}/(n_{\rm p}\kb T)$ is 
shown as a function of dimensionless time $t/\lambda_{\rm H}$ 
for equilibrium initial conditions and different strengths of the 
flow. 
The finite extensibility parameter was chosen to be $b=50$. 
Details of the numerical integration of Eq.~(\ref{dotA_susy}) 
and evaluation of $R_k$ and $\tau^{\rm p}$ with the distribution 
function (\ref{psi_QEAsusy}) are given in Sec.~\ref{DualInt}. 
As seen from Fig.~\ref{QEAsusy_fig} a), 
we observe that for weak or moderately strong flows, 
$\hat{\kappa} \lesssim 0.5$, the analogue of the conformation 
tensor ($n=1$) already provides a satisfactory description. 
Within the resolution of Fig.~\ref{QEAsusy_fig} a), 
the predictions are the same also for higher values of $n$. 
We observe that the description in terms of the second moment 
becomes worse in stronger flows, $\hat{\kappa} > 0.5$
Including higher order moments,  
we observe a fast convergence of the polymer stress in the 
macroscopic description to the result of the microscopic simulation. 
As seen in Fig.~\ref{QEAsusy_fig} b), the first three even moments 
already provide a satisfactory description for $\hat{\kappa}=1$. 

\begin{figure}[h]
        \setlength{\unitlength}{1cm}
        \begin{picture}(6,6)
        \put(-10,1){\centerline{\includegraphics[width=6cm]{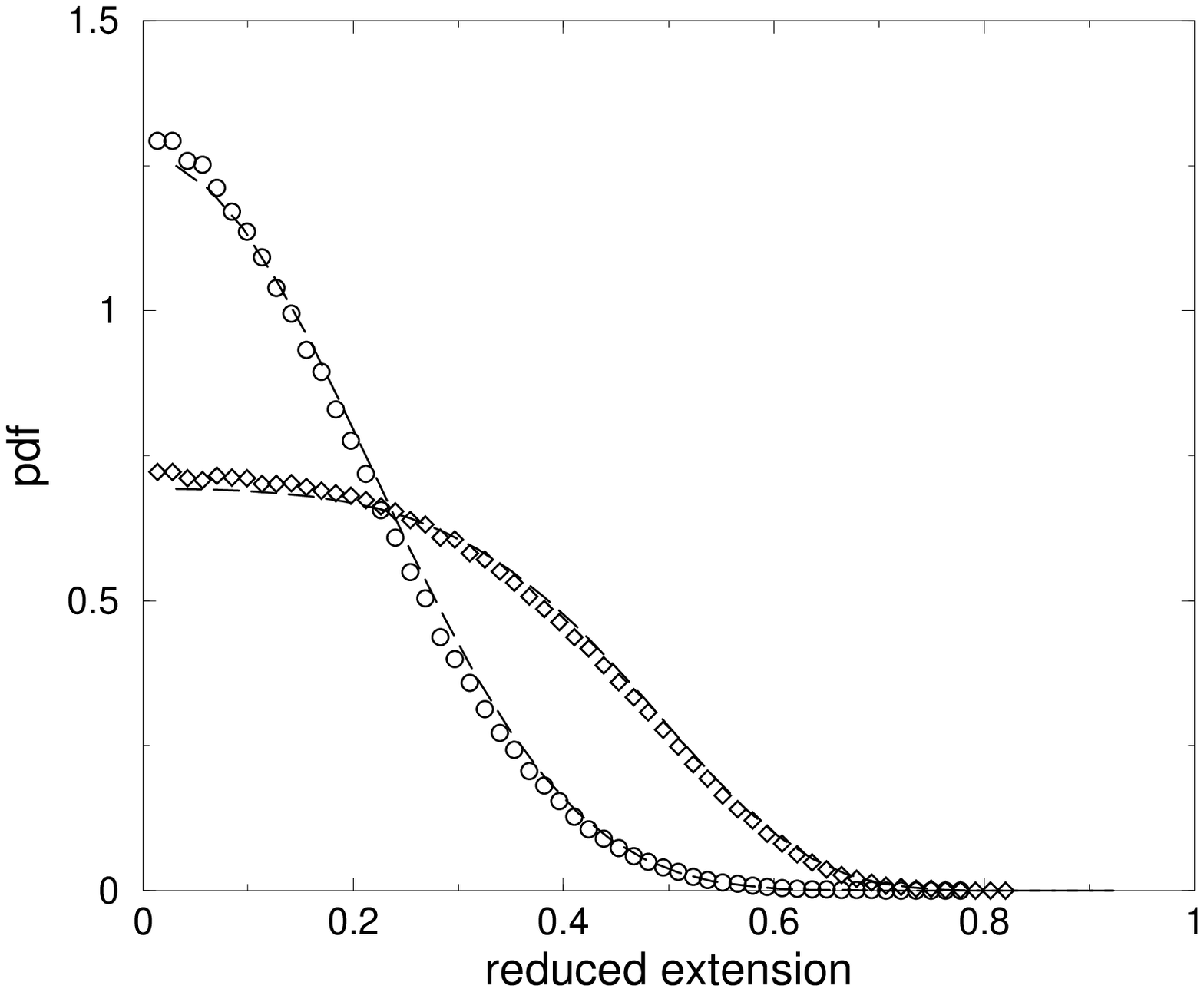}}}
        \put(-3,1){\centerline{\includegraphics[width=6cm]{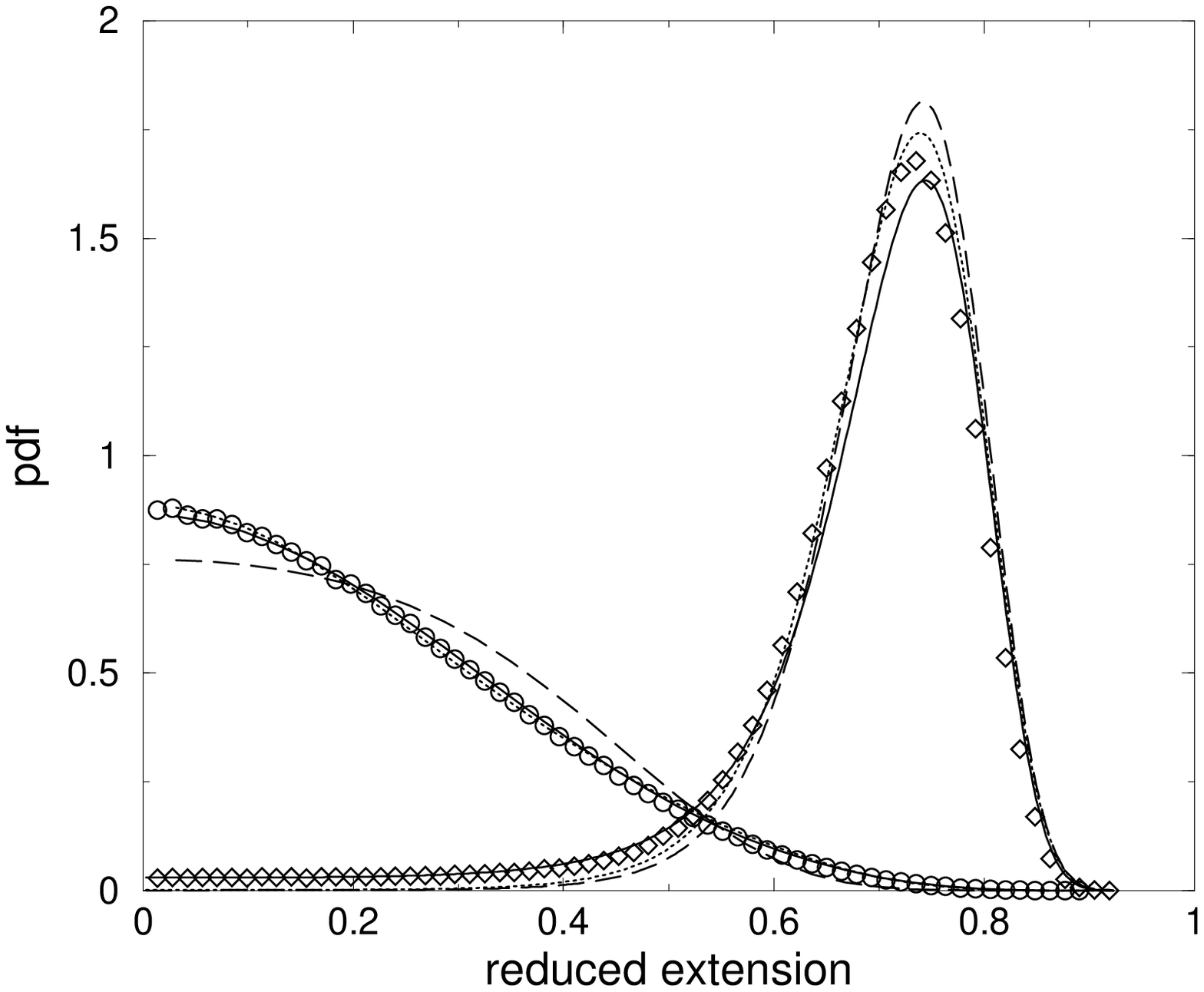}}}
\end{picture}
     \caption[QEA for SUSY] { \label{QEAdistr_fig} 
        Performance of QEA in steady elongational flow with 
        $\hat{\kappa}=0.5$ 
        $\mbox{\rm a)}$ and $\hat{\kappa}=1.0$ $\mbox{\rm b)}$. 
        Shown is the distribution function as a function of reduced 
        extension $Q/Q_{\rm max}$ for 
        different sets of macroscopic variables at two times. 
        Dashed: $n=1$, dotted: $n=2$, solid: $n=3$ and 
        symbols: Brownian dynamics simulations of full kinetic model.}
\end{figure}

It is interesting to trace back the difficulties 
of the macroscopic description in terms of second moments to the 
behavior of the probability distribution function (pdf) in these flows. 
Since the macroscopic description corresponds to the manifold of 
distribution functions (\ref{psi_QEAsusy}), 
direct comparison between microscopic and 
macroscopic description is possible also on the level of distribution 
functions. 
In Fig.~\ref{QEAdistr_fig}, the quasi--equilibrium states 
(\ref{psi_QEAsusy}) for the corresponding values of the 
Lagrange multipliers are compared to the numerical solutions of 
the kinetic equation at time $t=2\lambda_{\rm H}$ and 
$t=8\lambda_{\rm H}$. 
In the present case, the kinetic equation is not solved directly  
but the stochastic differential equation 
associated with the kinetic equation is solved numerically for a 
large ensemble of dumbbells \cite{OE96}. 
This method is well--established and is known as Brownian dynamics 
simulation. In this case, the distribution function is represented 
by a histogram over the lengths of the dumbbells of the ensemble. 
As seen from Fig.~\ref{QEAdistr_fig} a), the QEA for $n=1$ gives a good 
approximation even for the distribution function in case the 
elongational flow is not too strong, $\hat{\kappa}\lesssim 0.5$. 
In stronger flows, the shape of the distribution function is not 
accurately predicted by the quasi--equilibrium distribution 
(\ref{psi_QEAsusy}) for $n=1$, both, for rather early times (circles) 
and later times (diamonds). 
A much better approximation of the distribution function for earlier 
times is obtained for $n=2$. 
Note, that for later times a characteristic peak of the 
distribution function near the maximum extension occurs 
\cite{Keu97}, indicating that most dumbbells are stretched to 
almost their maximum length. 
Note, that the precise shape of that peak is not described perfectly 
by the quasi--equilibrium distributions even for $n=3$. The approximation 
is however reasonably close to give accurate predictions for the polymer 
stress as seen in Fig.~\ref{QEAsusy_fig}.

\begin{figure}[h]
        \setlength{\unitlength}{1cm}
        \begin{picture}(6,6)
        \put(-5,1){\centerline{\includegraphics[width=6cm]{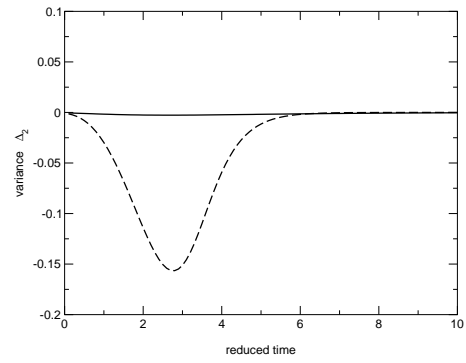}}}
\end{picture}
     \caption[QEA for SUSY] { \label{fig_defect} 
        Dynamic variance $\Delta_2$ as a function of reduced time 
        $t/\lambda_{\rm H}$. 
        Solid line corresponds to $\hat{\kappa}=0.5$, broken line to 
        $\hat{\kappa}=1.0$. 
        Comparison to Fig.~\ref{QEAsusy_fig} shows that the dynamic 
        variance $\Delta_2$ can be used as a measure for the accuracy 
        of the approximation.}
\end{figure}

As discussed in Sec.~\ref{defect}, the accuracy of the approximation 
of the dynamics by canonical distribution functions 
can be measured by a suitable norm of the dynamic variance $\Delta$. 
Let us reconsider the approximation (\ref{psi_QEAsusy}) with $n=1$. 
Then, the quantity $\Delta_2$ defined in Eq.~(\ref{Delta_ell}) 
for $\ell=2$ measures the difference of the microscopic and macroscopic 
dynamics of the fourth moment. 
For two strengths of the flow, $\Delta_2$ is given as a function 
of time in Fig.~\ref{fig_defect}.
From Fig.~\ref{fig_defect} one notices that $\Delta_2$ vanishes 
for early times $t/\lambda_{\rm H}\ll 1$ since the initial state 
was chosen to be the equilibrium distribution for which $\Delta_2=0$. 
Also for times $t\gg 1$, the dynamic variance $\Delta_2$ vanishes, 
since the steady state is correctly described by the approximation 
(\ref{psi_QEAsusy}). 
For intermediate times $t\approx 1$, $\Delta_2$ is different 
from zero and can be interpreted as a measure for the quality of 
the approximation. For $\hat{\kappa}=0.5$, $|\Delta_2|$ 
remains small, less than $10^{-3}$, while for $\hat{\kappa}=1.0$, 
$|\Delta_2|$ is two orders of magnitude larger. This is in agreement 
with the comparison to Brownian dynamics simulations shown in 
Fig.~\ref{QEAsusy_fig} that the approximation $n=1$ performs 
quite well for $\hat{\kappa}=0.5$, whereas it is much less reliable for 
$\hat{\kappa}=1.0$.
Note that $|\Delta_2|$ is a measure for the difference of the 
microscopic and macroscopic time evolution. Therefore, the maximum 
of $|\Delta_2|$ occurs at earlier times than the maximum 
difference of the moments and the polymer stress.

Although the model studied in this section is only one--dimensional, 
we expect the basic observations to be valid also in three 
dimensions. The present approach can easily be generalized to 
three dimensions and it would be interesting to confirm this 
conjecture. 
Remember that the quasi--equilibrium distributions (\ref{psi_QEAsusy}) 
are exact in equilibrium and in the steady state in case of potential flows. 
Therefore, it would be particularly interesting to see whether the 
above results also apply in case of non--potential flows in 
three dimensions. 

It is straightforward to apply the numerical integration scheme 
proposed in Sec.~\ref{DualInt} to the three--dimensional 
problem. 
The corresponding numerical integrations are however much more 
time consuming than in the one dimension. 
Under certain circumstances, these integrals may be evaluated 
efficiently, e.g.~by fast Fourier transform algorithms \cite{HuBa98}.

Based on the observations for this model, 
we conclude this section with the remark that the frequently used 
conformation tensor models might not perform very well in strong flows 
if compared to the original model. 
In the present example, the failure of the conformation tensor 
model is related to the so--called 
`coil--stretch transition' \cite{Ge74}, 
giving raise also to the instability of the invariant manifold 
as studied in \cite{IlKa00a}.

\section{Conclusion}
\label{end}
In the present work, we have shown how the quasi--equilibrium 
or maximum entropy approximation can be applied to kinetic models 
of polymer dynamics. As a result, canonical distribution functions 
and constitutive equations of polymer dynamics are obtained. 
A numerical implementation of the constitutive equations is proposed 
together with a measure for the accuracy of the quasi--equilibrium 
approximation. 
The numerical implementation and the measure of the accuracy 
of the approximation is illustrated in detail for an example of 
polymer dynamics.  
It is found that the accuracy of the approximation increases with 
increasing number of macroscopic variables. In steady elongational 
flow, more macroscopic variables are needed in order to arrive at the 
same accuracy above the coil--stretch transition than below. 

Finally, we mention that the same methods can be applied to other 
kinetic models of polymer dynamics like, e.g.~liquid--crystalline polymers 
\cite{IKO02}.

\clearpage

\clearpage
%
%

\clearpage
%
%

\end{document}